\documentclass{mem}
\usepackage{natbib}\usepackage{txfonts}\usepackage{balance}
\usepackage{graphicx}
\usepackage[a4paper]{hyperref}
\begin{document}
\def\teff{$T\rm_{eff }$}
\def\kms{$\mathrm {km s}^{-1}$}

\title{
Identification of important VO spectral services benefiting from deployment on the Grid
}

\subtitle{}

\author{
P. \,\v{S}koda} 


\institute{
Astronomical Institute of the Academy of Sciences,
Fri\v{c}ova 298,
251\,65\,Ond\v{r}ejov,\\
Czech Republic\\
\email{skoda@sunstel.asu.cas.cz}
}
\authorrunning{\v{S}koda }

\titlerunning{VO spectral services on the GRID}

\abstract{ The majority of VO-compatible spectra handling applications operates
only with a few spectra entirely downloaded from single or several SSAP
servers. We try to identify the scientific cases which could immediately
benefit from future SSAP applications designed for GRID deployment. Their key
feature is the sophisticated spectra pre-selection and preprocessing done on
distributed servers using the intelligent agent summarising the results and
performing final high-level processing or analysis. 
 }

\maketitle{}

\section{Introduction} 

There is a wealth of calibrated astronomical spectra accessible in current data
archives, however most of them are not suitable for direct physical analysis.
Being pre-reduced by some automatic pipeline or individually by manual
reduction, they  are mostly stored in archives in form of  FITS file or ASCII
table as the relation of intensity in arbitrary numbers (instrumental counts,
digital numbers etc.) and wavelength (or frequency).

The  scientific analysis of such  spectra requires further processing by the
variety of different methods. In certain studies a huge number of spectra has
to be collected from different servers (e.g. in different spectral regions )
and transformed into  common units. 

The clear definition of metadata description and easy unit conversion as well
as transparent  lookup and download of spectra in Virtual Observatory has a
great potential to become a new advanced way of astronomical research. The
connection of powerful distributed computing available by GRID, Web Services
and VO protocols can establish an innovative  research environment allowing the
data mining  of huge datasets.

\section{Current status of VO tools for spectroscopy}

Although there is a number of various tools written for the analysis of spectra
of astronomical objects (e.g IRAF packages {\tt splot} and {\tt
spectool}, Starlink {\tt DIPSO}, MIDAS {\tt XALICE} as well as numerous single-purpose C,
FORTRAN and IDL routines), the number of modern tools understanding the VO
protocols (especially spectra access protocol - SSAP) is still very low.  We
are giving a brief overview of their capabilities and status of  development 
(in July 2008).

\subsection{VOSpec}
{\small\tt http://esavo.esa.int/vospec/}
\begin{itemize}
\item Developed at ESA
\item Very simple (mainly for building SED) 
\item Polynomial and Gaussian fits only
\item Blackbody fit 
\item No RV measurement built-in 
\item It can be called directly from  VizieR
\item Can work with linelists through Simple Line Access Protocol (SLAP)
\item Theoretical VO (TVO) supported (synthetic spectra: Kurucz, disks)
\item Rapid development  
\item Integration of some methods in users own program possible
\item JAVA applet 
\item Contains PLASTIC VO-interoperability layer  
\item De-reddening of extragalactic objects built-in
\item Support of dimensional equations for units (DIMEQ, SCALEQ)
\item Rather complicated view of users data - needs to create SSA wrapper and
prepare VOTable even for one simple FITS spectrum before viewing 
\end{itemize}

\subsection{SpecView}
{\small\tt http://www.stsci.edu/resources/\\
software\_hardware/specview}
\begin{itemize}
\item It is available as JAVA applet or  stand-alone application
\item Developed and supported  by STScI
\item Understands a number of formats from HST instruments and most NASA/ESA satellites
\item Easy work with local data (binary or ASCII tables and 1D FITS)
\item Well  tailored for practical  spectral analysis
\item Powerful linelists (many included)
\item Fitting of models by  $\chi^2$ --- even user models may be constructed
\item Built-in number of standard stars spectra and models for various physical conditions
\item Whole Kurucz library of spectra available for immediate display.
\item Only simple polynomials for visual fit available
\item De-reddening, CLOUDY models built-in
\item Proper error propagation of fitted values 
\item Support of dimensional equations
\item Does not support PLASTIC
\end{itemize}

\subsection{SPLAT-VO}
{\small\tt http://star-www.dur.ac.uk/\textasciitilde pdraper/\\
 splat/splat-vo/}
\begin{itemize}
\item Supported by JCMT after closing Starlink
\item Most advanced for stellar astronomy (fitting by freehand drawing, INTEP)
\item RV measurement - both Gaussian fit and mirroring
\item Custom line list for individual spectra available
\item Wavelet analysis
\item Full featured data and editor and spreadsheet 
\item Publication quality output, powerful plotting options, annotations
\item Supported PLASTIC
\item Reads 1D FITS image files (e.g. rebinned IRAF multispec files) 
\item Asynchronous SSAP queries
\end{itemize}

\subsection{Period04}
{\small\tt http://www.univie.ac.at/tops/Period04/}
\begin{itemize}
\item Its predecessor was  widely used Period98
\item Period04 is rewritten in Java (and partly C) 
\item Needs formatted text files
\item supports VOTable files
\item Several period finding methods built-in
\item Can handle even the regular period shifts
\item PLASTIC layer under preparation
\end{itemize}

\subsection{FROG}
{\small\tt http://www.jach.hawaii.edu/software/\\
starlink/}
\begin{itemize}
\item Part of  Starlink Java package 
\item Included in last JCMT Starlink  release (Hokulei) 
\item VO-protocols built-in
\item Almost same capabilities as Period04
\item Not developed after closing Starlink
\item Last version since 2004
\item The period searching engine available remotely as a SOAP service
\item Web service for Fourier transform
\item Collaborates with TOPCAT, but not PLASTIC
\item Easy but powerful for doing period analysis in VO environment
\end{itemize}

\subsection{SDDS Spectrum Services}
{\small\tt http://voservices.net/spectrum}
\begin{itemize}
\item Advanced access and visualization of Sloan spectra 
\item Search of similar spectra using PCA method
\item Several continuum and line fitting procedures 
\item Programmable interface available but in .Net or Mono framework
\item Collaboration with other SSA servers not supported - Only SDSS data at JHU
\end{itemize}

\section{Spectra post-processing}

The current VO applications using SSA protocol are very simple providing only
the URL of the given dataset, so the client gets a whole (often quite large)
file. The further processing is then done fully by the client who has to
download all the relevant datasets and keep them in the memory or local
storage.  The real strength of the VO technology lies, however, in transferring
part of the client's work to the server, which usually runs on a powerful
machine with fast connection to the data archives.  The typical example of
commonly required post-processing of fully reduced spectra (at least in stellar
astronomy) are given below:

\begin{itemize} 
\item Cutout services (selection of only  certain spectral
lines or regions giving the wavelength range)  
\item Projection of multidimensional datasets (in 3D spectroscopy)
\item Rectification of continuum
\item Rebinning to given, usually equidistant grid of wavelengths (constant
$\Delta\lambda$ or $\Delta\ln\lambda$)
\item (De)convolution of instrumental  profile 
\item Application of physical broadening functions (rotation, limb darkening)
\item Shift in radial velocity, application of heliocentric correction computed
on server 
\item Merging of individual echelle orders in a single spectrum 
\end{itemize}

Although most of these post-processing methods can be implemented in a
straightforward way, the rectification of continua may be a difficult art or
black magic. Especially two problems are most common and most difficult to
solve easily.
\vspace{1ex}

\noindent {\it Normalising echelle spectra} \\
A really challenging problem (which had not yet been solved fully) is the
normalisation of echelle spectra - especially with wide absorption lines or
complicated profiles (P Cyg, Be stars). Some wide lines may span several
echelle orders and so the individual orders have to be precisely unblazed and
then merged. This problem is very complex, depending on construction of
individual spectrograph, the observation strategy etc. The explanation of
problem and partly the solution is given by \citet{2003adass..12..415S},
\citet{skosle:2004}, \citet{2001A&A...369.1048P} or
\citet{2002A&A...383..227E}.

\vspace{1ex}
\noindent {\it $\chi^2$ fitting procedures} \\
The more advanced way of normalising is the least squares fitting based on some
theoretical model. It is very popular in radioastronomy and X-ray astronomy (XSPEC) 
rather for estimation of the global SED fit parameters (usually only as
combination of several simple power law functions). Quite useful in optical
spectroscopy is the usage of appropriate synthetic spectrum to find the real
continuum in data. The implementation of such a procedure in VO tools could
call the Theory VO (TVO) servers to get correct models (e.g. Kurucz) and, if
necessary, to improve the models by recomputing on the GRID in a iterative way.
Unfortunately, the models of many interesting objects  are very speculative.

\section{Advanced spectral analysis}

As the goal of VO is to make easy and comfortable the physical analysis of a
huge number of fully reduced (and post-processed) spectra in the environment of
VO client (or web portal), the common recipes of spectral analysis have to be
implemented as VO-compatible.  While the simplest ones may be done by client,
the full power of VO may be exploited only by their implementation on server
side (and best deployed on GRID due to their inherent independencies).  Here
are the examples of most commonly used techniques of spectra analysis in the
stellar astronomy:

\subsection{Dynamical  spectrum } 

It is sometimes called the gray representation or trailed spectrum. The basic
idea is to find the small time-dependent deviations  of individual line
profiles from some average. 
\begin{figure} 

\centering 
\includegraphics[width=1.00\linewidth]{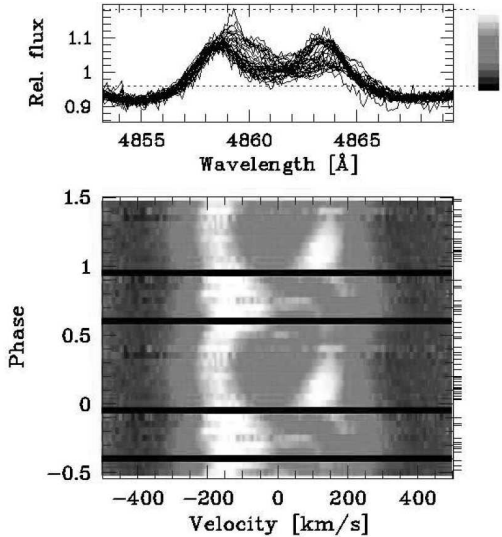}
\caption{Dynamical spectrum of H$\beta$  line profile variability of 59~Cyg.
Residuals from average profile of 38 spectra  are binned to 20 phase bins
corresponding to period 28.192 days. Two expanded cycles shown for clarity.
Individual profiles are overplotted above.  After \citet{maintz:2003}
\label{dynspec2}}
\end{figure}

First the average of many high dispersion high SNR spectra (with removal of
outliers) is prepared (called template spectrum). Then each individual spectrum
in time series is either divided by the template (quotient spectrum) or the
template is subtracted from it (the differential spectrum). The group of
similar resulting intensities is given the same colour or  level of gray.  See
Fig.~\ref{dynspec2}.\hspace{1em}
More examples may be found e.g.  in
\citet{1999A&A...345..172D}, \citet{maintz:2003} or \cite{uytterhoeven:2004}.

The result is drawn in 2D image where on horizontal axis is a wavelength in the
line profile or corresponding RV relative to laboratory wavelength, on the
vertical axis the time of middle of observation (in HJD) or the circular phase
when the data are folded with certain period.  

\subsection{Measurement of radial velocity and higher moments of line profile}

The one of the important information received from spectrum is the radial
velocity (RV) of the object. From its changes the binarity can be revealed, or the
possession of extrasolar planet. RV is sometimes presented  as zero-th moment
of the spectral line profile. The higher moments of the line profile are
important as well. The first moment is equivalent width. The combination of
higher moments of line profile is a one of the possible ways of determination
of non radial pulsation modes -- numbers $l,m$ \citep{1992A&A...266..294A}. The RV
is usually measured from single lines (by finding their centre using fits or
mirroring) or from part of spectrum by cross-correlation techniques using the
template spectra (observed or synthetic) of similar object.

\subsection{Measurement of equivalent width}

The Equivalent width (EW) of the spectral line gives the information about the
number of absorbing or emitting atoms of given element.  The measurement of
equivalent width (EW) of a given line needs the determination of area bordered
by line profile and the continuum. This area is transformed to the rectangular
line with the depth of 1.0. The width of this rectangle is called equivalent
width. Emission lines by definition have negative EW.  The changes in EW during
time may bring about the information about the dynamic evolution of the target
and may be subjected to period analysis.  

Due to its dependency on continuum placement, a problem of automatic
rectification arises again.  It is extremely difficult on echelle spectra and
spectra of late type stars.  If not rectified properly, the extremely shallow
lines may give large error in EW \citep{1997A&A...320..878V}.

EW are often used in abundance analysis. A good check of correct normalisation
is the comparison of EW of spectral lines of the same element in different
spectral regions \citep{2000A&A...358..553H}.

Some attempts of automatic rectification have been proposed eg. by
\citet{Zhao2006} but in general another kind of information (synthetic spectrum
or template from other instrument) is required.  

The measurement of EW of a number of lines  in thousands of spectra  may be
accomplished as a VO service and run in parallel on a GRID. 

\subsection{Bisector analysis}

\begin{figure} 
\centering 
\includegraphics[width=1.05\linewidth]{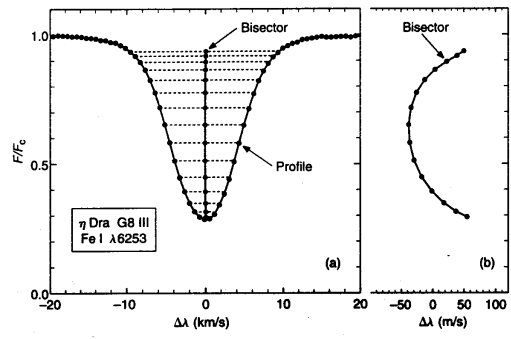}
\caption{The construction of bisector of line profile (left panel) and the zoomed bisector (right panel). 
From \citet{1982ApJ...255..200G}.
\label{bisector}}
\end{figure}

It is a method describing quantitatively the tiny asymmetry or subtle changes
in line profiles.  It is easily done by marking the middle of horizontal cuts
of profile in different line depths.  The line connecting such points (called
bisector) is then zoomed in horizontal (wavelength or RV) direction. See
Fig.~\ref{bisector}.

The characteristic shape of  bisector gives the information about turbulence
fields (e.g. convection) in stellar photosphere, characterised by the value of
micro-turbulent velocity \citep{1982ApJ...255..200G, 2005PASP..117..711G} or
about other processes causing the tiny profile asymmetry.  It has been used
successfully for searching of extrasolar planets \citep{2001AJ....121.1136P} or
in asteroseismology.  Requires high resolution  (echelle) and high SNR
normalised spectra. 

\subsection{Period Analysis}

One of the most popular methods used in astronomy is the period analysis.  Its
aim is to find the hidden periods of variability of given object.  Sometimes
this period can be identified with some physical mechanism (e.g. orbital period
of binaries, rotational modulation or pulsations). Wide range of objects show
the multi-periodicity on various time scales (e.g. binary with pulsating
components).  Ones the suspected period is found, the data may be folded
accordingly, plotted in circular phase corresponding this period. Very helpful
is a interactive capability of showing data folded while selecting different
peaks at the periodogram.

The current availability of VO tools is limited.  The nice FROG
(part of Starlink new Java-based development) is abandoned and so  the only
hope is the Period04 by P. Lenz, who is working on the PLASTIC compatible data
exchange of VO tables (Lenz, 2008, private communication).

\begin{figure}[ht!] 
\centering
\includegraphics[width=1.00\linewidth]{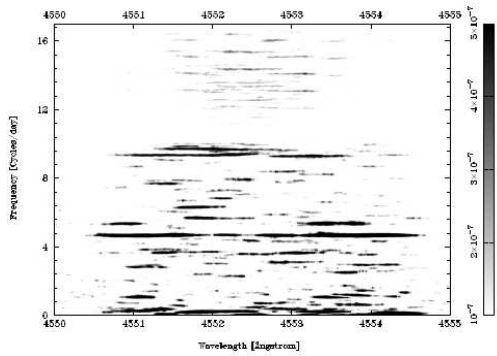}
\caption{Example of 2D periodogram of SiIII 4552~\AA\ line profile variability
of $\lambda$~Sco. 
The darker gray level marks the higher value of power
spectrum in given wavelength for given period. 
After \citet{uytterhoeven:2004}.
\label{dynperiod}}
\end{figure}

\noindent{\it Periodogram of line profiles}.
It is a kind of gray representation with
period on vertical axis and the zoomed line profile on horizontal. See
Fig.~\ref{dynperiod}.  Works well for analysis of multi-periodical non-radial
pulsations \citep{1999A&A...345..172D}.
\subsection{Doppler imaging }

It was introduced by \citet{1983PASP...95..565V} as a method allowing the
surface mapping of stellar spots.  First test were done on stars of RS CVn type
and on $\zeta$~Oph \citep{1983ApJ...275..661V}. Works well on rapid rotators
and needs a high resolution spectra with very high SNR (300--500). Tho whole
rotational period should be covered well, better several times.  When all the
requirements are met,  the map  of surface features (spots, nodes of non radial
pulsations) is obtained with very high accuracy. See the left panel of
Fig.~\ref{zeemandop}.
\subsection{Zeeman Doppler Imaging}

\begin{figure} 
\centering 
\includegraphics[width=1.00\linewidth]{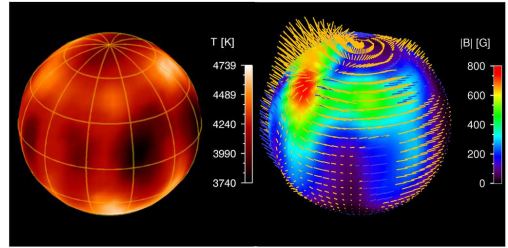}
\caption{Zeeman Doppler Imaging of II Peg 
After \citet{Strassmeier2008}
\label{zeemandop}}
\end{figure}

Quite complicated processing of spectra is required for study of stellar
magnetic fields.  The estimation of magnetic field from  polarimetry using the
Zeeman phenomena involves the processing of long series of homogeneous spectra
to be accomplished in parallel with extreme precision and requires again the
information from synthetic models (simulation of Stokes parameters on simulated
magnetic stars) The nice example is the model of II Peg by
\citet{Strassmeier2008}. See the right panel of Fig.~\ref{zeemandop}.
\subsection{Spectra disentangling}

This method allows to separate the spectra of individual stars in binary or
multiple systems and simultaneously to find orbital parameters of the system,
even in case of heavy blending of lines.  It supposes the changes in line
profile are caused only by combination of Doppler shifted components (no
intrinsic variability of star).  The best solution of orbital parameters and
disentangled line profiles of individual stellar components are found by least
square global minimisation.  The method also enables to  remove the telluric
lines with great precision.  The good orbital coverage and the estimate of
orbital parameters is required. Two approaches exist:  
\vspace{1ex}

\noindent {\it Wavelength space disentangling}\\
developed by \citet{1991ApJ...376..266B} and improved by
\citet{1994A&A...281..286S}.  It needs a large memory to store sparse matrices,
requires large computing power.  It is more straightforward to understand the
results and sources of errors.
\vspace{1ex}

\noindent {\it Fourier space disentangling} \\
introduced by \citet{1995A&AS..114..393H,1997A&AS..122..581H} in
program KOREL. Another program available today (still based on KOREL ideas) is
FDBINARY \citep{2004ASPC..318..111I}.  They  work in Fourier space, and
transform the wavelengths into $\ln\,\lambda$.  They solve a small amount of
linear equations, so they are   memory savvy and can be run on even small
computer.  The method, however,  requires  perfect continuum fit (difficult to
achieve at merged echelle spectra). On unreliably rectified spectra strange
artifacts may appear \citep{2004ASPC..318..111I}.

\subsection{Deconvolution of Iodine lines}

One of the new  approaches to high precision radial velocity measurement is the
usage of Iodine vapour cells in front of the spectrograph. The idea is to use
the huge amount of sharp Iodine lines as a ruler for marking exact position of
investigated stellar line thus eliminating the instrumental shifts. 

However, the Iodine lines have to be first removed from spectra to see the
stellar ones.  The process is not so straightforward and requires the models
and templates from high resolution Fourier Transform Spectrographs as well as
exposures without the cell. Finally the global optimization (e.g. by maximum
entropy method) is required.  See e.g. \citet{Fiorenzano2006}.

\subsection{Classification of stellar spectra}

It is a one of the methods where the full power of the Virtual observatory
infrastructure could be exploited to work on massive data volumes.  The goal is
to assign to every spectrum a short string  of letters and numbers representing
the physical characteristics of the star (surface temperature,  size, mass,
peculiar behaviour etc.).

It used to be a time consuming method done by visual comparison of many spectra from
photographic plates (e.g. MKK classification).  Today it has been done finding
the minimal differences between a grid of template (or even synthetic) spectra
and the examined one.  The techniques of artificial intelligence (e.g. neural
networks, genetic algorithms, self-organizing maps, cluster analysis techniques) 
may be very efficient, but the global optimization using the $\chi^2$
minimization seems to be more reliable with possibility of iterative control.
An example of automatic classification engine for white dwarfs is  described 
by \citet{Winter2006}.

As it can be seen, the algorithm can run in parallel and uses workflows, so it
may be nicely implemented as Grid application of VO services.

\section{Spectral analysis in Grid supported VO services}

\subsection{Workflows and pipelines}

Many astronomical legacy applications are designed to run as a single task on
local PC, the exchange of results with other tools is done over the local files
and a processing is driven by a set of parameters.  Some packages have to be
run as separate tasks in given order as some task are dependent on  a
information written to the  processed file by the preceding tasks.

So the work has to be  done in a strictly serial manner --- step by step.  Thus
the idea of workflows is quite natural for the most of scientist's work but
there is a strong reluctance of astronomers (driven by a strong conservatism
and habits) to use modern workflow builders/executors in the data reduction or
spectra analysis.  The certain  level of workflow management is accepted  for
distributed computing of models on clusters --- usually using batch queue
scheduler scripts either on GRID or on institutional network of PC (e.g.
Condor, Beowulf).

\subsection{Massive parallel spectra reduction}

The processing of spectra coming from current sophisticated spectrographs is a
quite computer demanding task and thus may be the good candidate for processing
on GRID.  Especially the rigorously correct reduction of echelle spectra
requiring the non-linear Least Squares optimization \citep{2002A&A...385.1095P}
or even the estimate of background light model \citep{hamscat} for the hundreds
of echelle exposures would benefit from GRID as it can be run in parallel for
all frames using the same calibrations (flat, bias) and templates (order
position, aperture estimates).  Some pipeplines are already written using the
concept of  workflows  (e.g. Gasgano, Reflex for ESO UVES and FORS
spectrograph, respectively).

\subsection{Advantages of  VO services} 

The key VO feature, that allows the novel approach to the everyday scientific
work and thus promises new discoveries, is the interoperability of all
VO-compatible tools and common format for describing metadata (VOTable).

It allows (in principle) the most of current work of an astronomer to be done
in one GUI which can provide an transparent access to VO tools accomplishing
all of the most tedious and routine work: 

The VO infrastructure enables transparent  search, download and on-the-fly
conversion of data, unified presentation of  different relations from
multi-dimensional datasets, the complicated reduction and analysis tasks to be
done in background or deployed on GRID, the results may be stored  directly
in databases or staged on virtual network home space (like MySpace).  The most
sophisticated data processing may be implemented in the form of  web services
(WS).

Although most of the methods described may be (and have been) run in local
environment using the legacy astronomical application, their implementation as
VO services would benefit from the VO infrastructure and might allow the
analysis of a huge amount of data easily.  The key issues in VO approach are
the 
\begin{itemize} 
\item Unified data format (VO-Table, semantics of
variables) 
\item Quick search for available data in a global manner using SSAP
and registries 
\item Transparent data conversion, homogenization, rescaling
\item Powerful presentation with remote data and  theoretical  results (TVO)
\item Staging disks with virtual home (VOSpace, MySpace)
\end{itemize}
Combining the VO infrastructure power and the easy and transparent high
performance computing on GRID will allow the advanced analysis of  large
spectral surveys  feasible in a reasonable time. It may force even the kind of
serendipitous research --- e.g. you may  click on star in the  image of
globular cluster to see its dynamic spectra folded with estimated period.

\section{Killer spectral applications for VO}

Here we give some scientific cases that may be the killer applications  forcing
the stellar astronomers to use the VO tools suggested above despite their
current reluctance and hesitation:
\begin{itemize} 
\item Use VO to find all stars with emission in given line, find the time when
it was in emission and plot the time evolution of its EW.

\item Use VO to get 1000 or more spectra of the given object, cut out regions
around given lines, plot the lines, make a gray dynamic spectrum folded in time

\item The same, but fold by period clicked on interactive periodogram

\item Get the unknown lines identification of piece of spectra from theoretical
observatory (Simple line access protocol) having the line selection limited by
pre-estimated temperature (using Saha equation)

\item Create light and radial velocity curve of a binary star for given period
(estimated by other VO tool running in parallel and exchanging data over
PLASTIC)

\item Fit the grid of models ($T_{\rm eff}, \log\, g$) to the observed spectrum
for many stars (e.g. from given cluster) 
\end{itemize}

\section{Conclusions}

Astronomical spectroscopy uses a wide range of techniques with different
level of complexity to achieve its final goal --- to  estimate the most
precise and reliable  information about celestial objects. The  large part
of spectroscopic analysis today  has been accomplished by several
independent non VO-compatible legacy packages, where  each works  with
different local files in its own data format. Analysis of large number of
spectra is thus very tedious work requiring good data bookkeeping. 

Accomplishing the analysis in VO infrastructure may benefit from automatic
aggregation of distributed archive resources (e.g. the multispectral
research), seamless on-the-fly data conversion, common interoperability of
all tools  (using PLASTIC protocol) and powerful graphical visualisation of
measured and derived quantities. 

The currently available VO clients supporting the SSA protocol can provide with
only basic functions in a interactive environment.  The more advanced work with
spectra (e.g. automatic preprocessing or advanced analysis) is not supported in
VO at all.  The most of the legacy applications should be turned into the VO
server side services and the conversion of legacy scripts and recipes into
Workflows will allow their easy deployment on GRIDs.  

Modern techniques of astronomical analysis require the considerable amount of
computing power and very offten the iterative comparison with theoretical
models is indispensable.  The properly designed VO service getting both
observed and synthetic spectra transparently from VO resources and processing
the number of objects in parallel on GRID might become a killer application
turning the attention of wide astronomical community to the Virtual observatory
as a viable and innovative way of modern astronomical research.

By introduction of  modern  VO-aware tools into the  astronomical spectral
analysis  a remarkable increase of effectiveness of astronomical research can
be achieved.

\begin{acknowledgements}
This work has been supported by grant GACR 205/06/0584 and EURO-VO DCA WP6.
The Astronomical Institute Ond\v{r}ejov is supported by project AV0Z10030501
\end{acknowledgements}

\bibliographystyle{aa}

\end{document}